\documentclass[aps,prd]{revtex4}
\usepackage{graphicx}
\begin{document}

\title{ Estimation of Primordial Spectrum with post-WMAP 3 year data}
\author{Arman Shafieloo and  Tarun Souradeep}
\affiliation{Inter-University Centre for Astronomy and Astrophysics (IUCAA),
Ganeshkhind, Pune-411007, India}

\date{September 2007}

%\date{}

%%%%%%%%%%%%%%%%%%%%%%%%%% GENERAL MACROS USED %%%%%%%%%%%%%%%%%%%%%%%%%%%
\def\be{\begin{equation}}
  \def\ee{\end{equation}}
\def\bea{\begin{eqnarray}}
  \def\eea{\end{eqnarray}}
\def\ie{{\it i.e.}\ }
\def\eg{{\it e.g.}\ }

\begin{abstract}

In this paper  we implement an improved (error sensitive)
Richardson-Lucy deconvolution algorithm on the measured angular
power spectrum from the WMAP 3 year data to determine the
primordial power spectrum assuming different points in the
 cosmological parameter space for a flat $\Lambda$CDM cosmological model.
We also present the preliminary results of the cosmological parameter estimation by assuming a free form of the primordial spectrum, for a reasonably large volume of the parameter space.
The recovered spectrum for a considerably large number of the points in the cosmological parameter space has a likelihood far better than a `best fit'
power law spectrum up to $\Delta \chi^2_{\rm eff} \approx -30$. We use Discrete 
Wavelet Transform (DWT) for smoothing the raw recovered spectrum from the
 binned data. The results obtained here reconfirm and sharpen the conclusion drawn from our previous analysis of the WMAP 1st year data. A sharp cut off around the horizon scale and a bump after the horizon scale seem to be a common feature for all of these reconstructed primordial spectra. We have shown that although the WMAP 3 year data prefers a lower value of matter density for a power law form of the primordial spectrum, for a free form of the spectrum, we can get a very good likelihood to the data for higher values of matter density. We have also shown that even a flat CDM model, allowing a free form of the primordial spectrum, can give a very high likelihood fit to the data. Theoretical interpretation of the results is open to the cosmology community. However, this work provides strong evidence that the data retains discriminatory power in the cosmological parameter space even when there is full freedom in choosing the primordial spectrum.
\end{abstract}
\maketitle

%\pacs{PACS number(s): 04.50.+h, 98.80.Hw}

\section{Introduction}
\label{sec:intro}
Increasingly accurate measurements of the
anisotropy in the temperature of the cosmic microwave background
(CMB) has ushered in an era of precision cosmology. A golden
decade of CMB anisotropy measurements by numerous experiments was
topped by the results from the data obtained by the Wilkinson
Microwave Anisotropy Probe (WMAP) ~\cite{ben_wmap03,sper_wmap06,hinshaw_wmap06}.
Under simple hypotheses for the spectrum of primordial
perturbations, exquisite estimates of the cosmological parameters
have been obtained from the angular power spectrum measurement by
WMAP data combined with other cosmological 
observations~\cite{sper_wmap03,sper_wmap06}. Precision measurements of
anisotropies in the cosmic microwave background, and also of the
clustering of large scale structure, suggest that the primordial
density perturbation is dominantly adiabatic and has a nearly
scale invariant spectrum~\cite{sel04,sper_wmap06}. This is in good
agreement with most simple inflationary scenarios which predict
power law or scale invariant forms of the primordial perturbation~\cite{inflation1,inflation2,inflation3}.
The data have also been used widely to put constraints on
different parametric forms of primordial spectrum, mostly
motivated by inflation~\cite{hiranya06,covi06,bridges06,jerome06,hamann,ringeval}.
However, despite the strong theoretical appeal and simplicity of a
featureless primordial spectrum, it is important to determine the
shape of the primordial power spectrum directly from observations
with minimal theoretical bias. Many model independent searches
have been made to look for features in the CMB primordial power
spectrum~\cite{bridle03,hanne04,pia03,pia05,sam06}. Accurate
measurements of the angular power spectrum over a wide range of
multipoles by WMAP has opened up the possibility of deconvolving
the primordial power spectrum for a given set of cosmological
parameters~\cite{max_zal02,mat_sas0203,prd04,bump05,prd07,kog03,kog05}.
Theoretically motivated models that give features in the power
spectrum have also been studied and compared in post-WMAP
literature~\cite{contaldi03,manoj03,sarkar04,sarkar05,sin_sour06,cline1,cline2,minu}.

The angular power spectrum, $C_{\ell}$, is a convolution of the primordial
power spectrum $P(k)$ generated in the early universe with a
radiative transport kernel, $G(l,k)$, that is determined by the
current values of the cosmological parameters indicated by other cosmological observations.  The remarkably precise observations of the angular power 
spectrum $C_l$ by WMAP, and the concordance of cosmological parameters 
measured from different cosmological observations opens up the avenue to
directly recover the initial power spectrum of the density
perturbation from the observations. The error-sensitive Richardson-Lucy (RL) 
method of deconvolution was shown to be a promising and effective method 
to recover the power spectrum of primordial  perturbations from the 
CMB angular power spectrum~\cite{prd04}. We have improved the deconvolution method by factoring out the normalization factors from the iteratively recovered primordial spectrum, $P(k)$, in the algorithm to remove the artifacts that were present at the two ends of the recovered spectrum in our previous work(corrected by template subtraction). For a given set of cosmological 
parameters, this method obtains the primordial power spectrum that `maximizes'
the likelihood to data.

In this paper we apply the method to the CMB
anisotropy spectrum given by WMAP 3 year data. We employ Discrete
Wavelet Transform (DWT) for smoothing the raw recovered spectrum
from the binned data. In this work we first present detailed results of an automated computation of the primordial power spectrum for $6$ distinct
points in the cosmological parameter space for flat $\Lambda$CDM
models using WMAP 3 year data. Each of these $6$ points in the parameter 
space has specific characteristics of interest. We also present the preliminary results of the cosmological parameter estimation optimized over the form of the primordial spectrum in a coarsely sampled volume of the parameter space. In this case, instead of
simply computing the likelihood for a given model of initial power
spectrum, one obtains the initial power spectrum that maximizes
the likelihood at a point and assigns that likelihood to that
point in the space of cosmological parameters. However our results for the cosmological parameter estimation, have a coarse resolution in spacing of the parameters and is also limited in volume of the parameter space covered. In principle it is possible to extend this work to explore the ``entire'' space of cosmological parameters with high resolution along the lines being done routinely. 

%At this point, computational costs are still high and beyond which we can 
%do the proper exploration in the whole parameter space easily. We defer the wide exploration of the cosmological parameter space to future work.

In Sections II we review in brief the Richardson-Lucy
deconvolution method and the improvements that we have made to adapt the method to our problem. In section III we explain the smoothing by discrete wavelet transform(DWT). The recovered spectrum from WMAP 3 year data for $6$ different points in the cosmological parameter space is described in Section IV. In Section V we discuss about cosmological parameter estimation by optimizing (maximizing) the likelihood over a free form of the primordial spectrum. Finally, in Section VI we discuss our results along with concluding remarks.

\section{Richardson-Lucy deconvolution Method}

The Richardson-Lucy (RL) algorithm was developed and is widely used in
the context of image reconstruction in
astronomy~\cite{lucy74,rich72}. However, the method has also been
successfully used in cosmology, to deproject the $3$-D correlation
function and power spectrum from the measured $2$-D angular
correlation and $2$-D power spectrum~\cite{baug_efs93,baug_efs94}.

The angular power spectrum, $C_l$, is a convolution of the initial power spectrum $P(k)$ generated in the early universe with a radiative transport kernel, $G(l,k)$, that is determined by the values of the cosmological parameters. In our application, we solve the inverse problem of determining
the primordial power spectrum, $P(k)$, from the measured angular power
spectrum, $C_l$, using the relation

\be 
C_l\,= \sum_i G(l,k_i )\,P(k_i).
\label{clsum}
\ee

In the above equation, the {\em `target'} measured angular power
spectrum, $C_l\equiv C_l^D$, is the data given by observations,
and the radiative transport kernel, \be G(l,k_i) = \frac{\Delta
k_i}{k_i}\,|{\Delta_{Tl}(k_i,\eta_0)}|^2\,, \label{glk} \ee
encodes the response of the present multipoles of the CMB
perturbed photon distribution function $\Delta_{Tl}(k_i,\eta_0)$
to unit of power per logarithm interval of wavenumber, $k$, in the
primordial perturbation spectrum. The kernel $G(l,k)$ is
completely fixed by the cosmological parameters of the {\em
`base'} cosmological model. The kernel $G(l,k)$ also includes the
effect of geometrical projection from the three dimensional
wavenumber, $k$, to the harmonic multipole, $l$ on the two
dimensional sphere.

Obtaining $P(k)$ from the measured $C_l$, for a given
$G(l,k)$, is clearly a deconvolution problem.  An important feature of
the problem is that $C_l^D$, $G(l,k)$ and $P(k)$ are all positive
definite. We employ an improved RL method to solve the inverse problem for
$P(k)$ in Eq.~(\ref{clsum}).  The advantage of RL method is that
positivity of the recovered $P(k)$ is automatically ensured, given
$G(l,k)$ is positive definite and $C_l$'s are positive. The RL
method, readily derived from elementary probability theory of distributions~\cite{lucy74}, is an iterative method that can
be neatly encoded into a simple recurrence relation. The power
spectrum $P^{(i+1)}(k)$ recovered at the iteration $(i+1)$ is given
by

\be
P^{(i+1)}(k)- P^{(i)}(k)\,=\, P^{(i)}(k)\,\sum_l\,\tilde G(l,k)\,\frac {\tilde C^D_l\,-C_l^{(i)}}{C_l^{(i)}}
\label{RLstd}
\ee
where $\tilde G(l,k)$ is the normalized kernel (on the $l$ space for all $k$ wavenumbers), $\tilde C^D_l$ is the normalized measured data (target) and $C_l^{(i)}$ is the angular power spectrum at the $i^{\rm th}$ iteration obtained from $C_l^{(i)}\,= \sum \tilde G(l,k)\,P^{(i)}(k)$ using the recovered power spectrum $P^{(i)}(k)$. It is important to remind the reader that due to the  formulation in terms of conditional probability distributions, the RL method requires the kernel, $G(l,k)$, data, $C_l$, and the target vector, $P(k)$, all to be normalized at the beginning,

\be \,\sum_l\, \tilde C_l\,=\,1; \quad \,\sum_k\, \tilde P^{(1)}(k)\,=\,1;
\quad \,\sum_l\, \tilde G(l,k)\,=\,1\,,  \ee 
where $\tilde P^{(1)}(k)$ is the normalized initial guess model of the primordial spectrum. The normalization factor, $\zeta_k \,=\, \sum_l\, G(l,k)\,$, should be factored out from the final form of the recovered primordial spectrum at the end of the procedure, 
\be
P(k)= P^{(n)}(k) / \zeta_k,
\ee
where $P^{(n)}(k)$ is the reconstructed primordial power spectrum at the last iteration and $P(k)$ is the final result. The  Eq.~(\ref{RLstd}) and Eq.~(\ref{clsum}) together summarizes the standard RL deconvolution method for obtaining the primordial power spectrum $P(k)$. The final recovered power spectrum is independent of the initial guess $P^{(1)}(k)$ (see discussion in Appendix of ref.~\cite{prd04}).

Due to noise and sample variance, the data $C_{\ell}^D$ is measured within
some non-zero error bars $\sigma_l$. The standard RL method does not
incorporate the error information and tends to iterate to fit
features of the noise, as well.  In our problem, this manifests
itself as a non-smooth deconvolved spectrum $P(k)$ from the binned
data that has poor likelihood with the full WMAP spectrum data.  We
devise a novel procedure  to make the RL method sensitive to the errors
$\sigma_l$ by modifying Eq.~(\ref{RLstd}) to

\be 
 P^{(i+1)}(k)-  P^{(i)}(k)\,=\, P^{(i)}(k)\,\sum_l\,\tilde G(l,k)\,\frac {\tilde C_l^D\,-C_l^{(i)}}{C_l^{(i)}}\,\,\,\tanh^2\,
\left[\frac{(\tilde C^D_l\,-C_l^{(i)})^2}{{\tilde \sigma_l}^2}\right].
\label{RLerr}
\ee 
The idea is to employ a `convergence' function to progressively
weigh down the contribution to the correction $P^{(i+1)}-P^{(i)}$ from
a multipole bin when $C_{\ell}^{(i)}$ iterate to a value close to $C_{\ell}^D$ within the error bar $\sigma_l$. This innovation significantly improves the WMAP
likelihood of the deconvolved spectrum. Further improvement of the likelihood to the full $C_{\ell}$ data is obtained by a subsequent step of wavelet based smoothing described in the next section. 
In this paper we have also improved the IRL method that we used in our previous paper to directly remove the artifacts at the two tails of the spectrum. Compared to our earlier implementation, we have eliminated the need to subtract a known model from the recovered spectrum to remove the artifacts at the two ends of the spectrum. We have achieved this by a slight change in the normalization procedure in the method, that corrects for the effect of very low amplitude of $G(l,k)$ at very small and very large $k$. In the previous paper, the normalization factor $\zeta_k$, was hidden in the finally deconvolved $P(k)$ and it was factored out at the end of the process. As we are dealing with an iterative process, every small artifact will affect the higher iterations more strongly. To avoid this, we separate the normalization factor $\zeta_k$ from the $P(k)$ from the beginning. This small modification gives rise to a big improvement in the method where the final form of the recovered spectrum is free from the artifacts at the two ends. Thus we do not need to use any template to remove these artifacts (as in our previous paper ~\cite{prd04}). So in our revised iterative process we have modified Eq.~(\ref{RLerr}) to

\be 
P^{(i+1)}(k)- P^{(i)}(k)\,=\,P^{(i)}(k)\,\sum_l\,\tilde G(l,k)\, \zeta_k \,\frac {\tilde C_l^D\,-C_l^{(i)}}{C_l^{(i)}}\,\,\,\tanh^2\,
\left[\frac{(\tilde C^D_l\,-C_l^{(i)})^2}{{\tilde \sigma_l}^2}\right],
\label{RLerr2} 
\ee
where the normalization factor, $\zeta_k$, is explicitly present in the main iterative equation and we do not need to remove it from the final form of the $P(k)$ at the end of the process,

\be
P(k)=  P^{(n)}(k).
\ee
The final form of the recovered spectrum was obtained after smoothing the spectrum by using Discrete Wavelet Transform as discussed in the following section.  

% after normalization of all the components, we generated the $P(k)$ in the iterative process where
%In fact the normalization factors were removed from the form of the $P(k)$ and was multiplied to the kernel at the beginning which means we have neglected the normalization of $G(l,k)$.

\section{Discrete Wavelet Transform}

Wavelet transforms provide a powerful tool for the analysis of
transient and non-stationary data and is particularly useful in
picking out characteristic variations at different resolutions or
scales. This linear transform separates a data set in the form of
%low-pass or average coefficients, resembling the data itself, and
low-pass or average coefficients, which reflect the average behavior
of the data, and wavelet or high-pass coefficients at different
levels, which capture the variations at corresponding scales. As
compared to Fourier or window Fourier transform, wavelets allow
optimal ``time-frequency'' localization in the real, as well as,
Fourier domain. The vocabulary of DWT stems from applications in one
dimensional time-stream signal trains, but has found wide application
in signal in other domains and dimensions. Specifically in our case,
the `signal' being transformed is the power spectrum, $P(k)$, a one dimensional function of wavenumber, $k$.

Wavelets are an orthonormal basis of small waves, with their
variations primarily concentrated in a finite region, which make them
ideal for analyzing localized `transient' signals. Wavelets can be
continuous or discrete. In the latter case, the basis elements are
strictly finite in size, enabling them to achieve localization, while
disentangling characteristic variations at different frequencies
\cite{Daubechies}. This is the preliminary reason for us to employ
discrete wavelets for our analysis. For more details about DWT and its 
theoretical basis, see \cite{prd07}.

In this paper, we use DWT to smooth the raw recovered spectrum obtained from the deconvolution using binned CMB spectrum data. The raw deconvolved spectrum has spurious
oscillations and features arising largely due to the $k$ space
sampling and binning in $l$ space. The main goal is to reconstruct the primordial spectrum which lead to an angular power spectrum with a high likelihood to the entire $C_{\ell}$ data at each multipole including the covariance between them. The WMAP likelihood of the $C_{\ell}$ suffers owing to the spurious oscillations in the spectra on scales smaller than $\ell$ multipole space bins.  The WMAP likelihood improves as the spectra, $P(k)$ is smoothed. We use DWT to smooth the recovered spectrum so that WMAP likelihood of the corresponding theoretical $C_l$ is maximized. Strictly speaking, the best likelihood obtained in our method is a lower bound leaving open a mathematical possibility of obtaining a superior likelihood at the given point of the cosmological parameter space with a different primordial power spectrum \footnote {There are some indications that the RL method can be related to a zero-noise limit of ML estimation.}.
Although it is difficult to establish that the final result is the unique solution with maximum likelihood, but numerous variations we have explored does suggest that it is perhaps very close to the best possible result. So we can claim that the improved reconstructed likelihoods which we drove for different points in the parameter space by assuming a broken scale invariant form of the primordial spectrum, put an upper limits for the best possible results.

For our smoothing purpose, we use DWT in a systematic way to separate the features of the raw recovered spectrum. We map the raw recovered spectrum which has 1400 discrete points to an array of $2^{11}$ points by padding at the two ends. By applying the discrete wavelet transform to the input file, we get corresponding $2^{11}$ wavelet coefficients. If we apply the inverse wavelet transform to whole set of derived wavelet coefficients, the orthonormality of the wavelet basis will lead back to the raw recovered spectrum. But if we exclude wavelet coefficients above a given level of resolution, then the inverse wavelet transform, leads to a smoothed power spectrum compared to raw deconvolved spectrum essentially filtering out spurious high resolution structures arising due to numerical 
effects ~\footnote{It is important to note that the deconvolution is based
on the binned $C_{\ell}$ angular power spectrum, implying a finite resolution in multipole (2D wavenumber) space. Hence is clear from information 
theoretic/physical intuition that all structures in the (3-D) power spectrum on scales smaller than the corresponding size of the wavenumber bin are bound to 
be spurious.}.

\begin{figure}[h]
  \includegraphics[scale=0.5, angle=0]{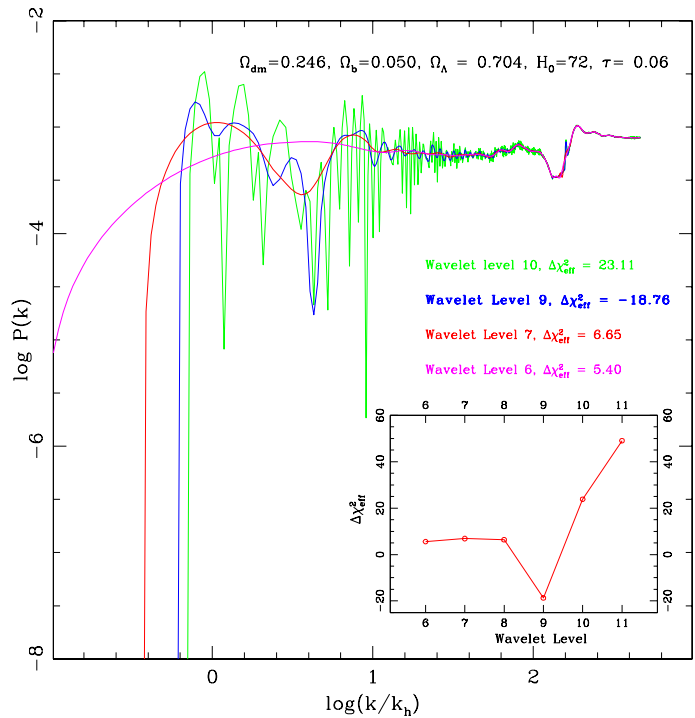}
  \caption{ Resultant $P(k)$ for a sample point in the cosmological parameter space is shown in the blue curve. The other curves show the P(k) recovered at different levels of DWT smoothing. The blue line which is the reconstructed result obtained by retaining all wavelet coefficients up to the 9th wavelet level has the best likelihood with $\Delta {\chi_{\rm eff}^2} = -18.76$ with respect to the best fit power-law primordial spectrum in the whole parameter space. We used $H_0=72, \Omega_{dm}=0.246, \Omega_b=0.05, \Omega_{\Lambda}=0.704, \tau=0.06$ as the cosmological parameters. The plot in the inset shows the resultant $\Delta {\chi_{\rm eff}^2}$ of the reconstructed results at different wavelet levels. The `optimality' of the $n=9$ level DWT smoothing in this case is clearly demonstrated.}
\label{fig_dwt}
\end{figure}

At the first step, we use the original raw recovered spectrum to calculate the likelihood to the WMAP-3 data. It is in fact the inverse wavelet transform of the whole wavelet coefficients which has $2^n, n=11$ coefficients. We call it the recovered spectrum at level 11. In the next step, we cut half of the coefficients and we use only the first $2^{10}$ coefficients. The derived results (recovered spectrum at level 10) would be smoother than the original spectrum. We continue the procedure for all different levels (n=2 to 11), and at each level we calculate the likelihood of the recovered results. The recovered result at a level which gives the best likelihood would be our final result. Using discrete wavelet transforms has this important advantage of being a well defined smoothing procedure that also retains identify localized features in P(k) that contribute a significant improvement to the likelihood. Figure \ref{fig_dwt} shows the reconstructed $P(k)$ for a sample point in the cosmological parameter space, smoothed up to different wavelet levels. The inset of the likelihood as a function of smoothing resolution shows that there exists an `optimal' level of smoothing which `maximizes' the likelihood with respect to data. The blue line in figure \ref{fig_dwt} is related to the 9th wavelet level where the best likelihood to the data is achieved. We reiterate that this allow the entire procedure to be automated so that an ``optimal'' primordial spectrum is recovered given the cosmological parameters.
   
%  Red line which has the highest distortion, is related to the 10th wavelet level. Thick green line is related to the 9th wavelet level which also has the best likelihood to the data. The other two lines are related to 7th and 6th wavelet levels. As we see at some specific wavelet level we can get the best likelihood.
 
\begin{figure}[h]
  \includegraphics[scale=0.8, angle=0]{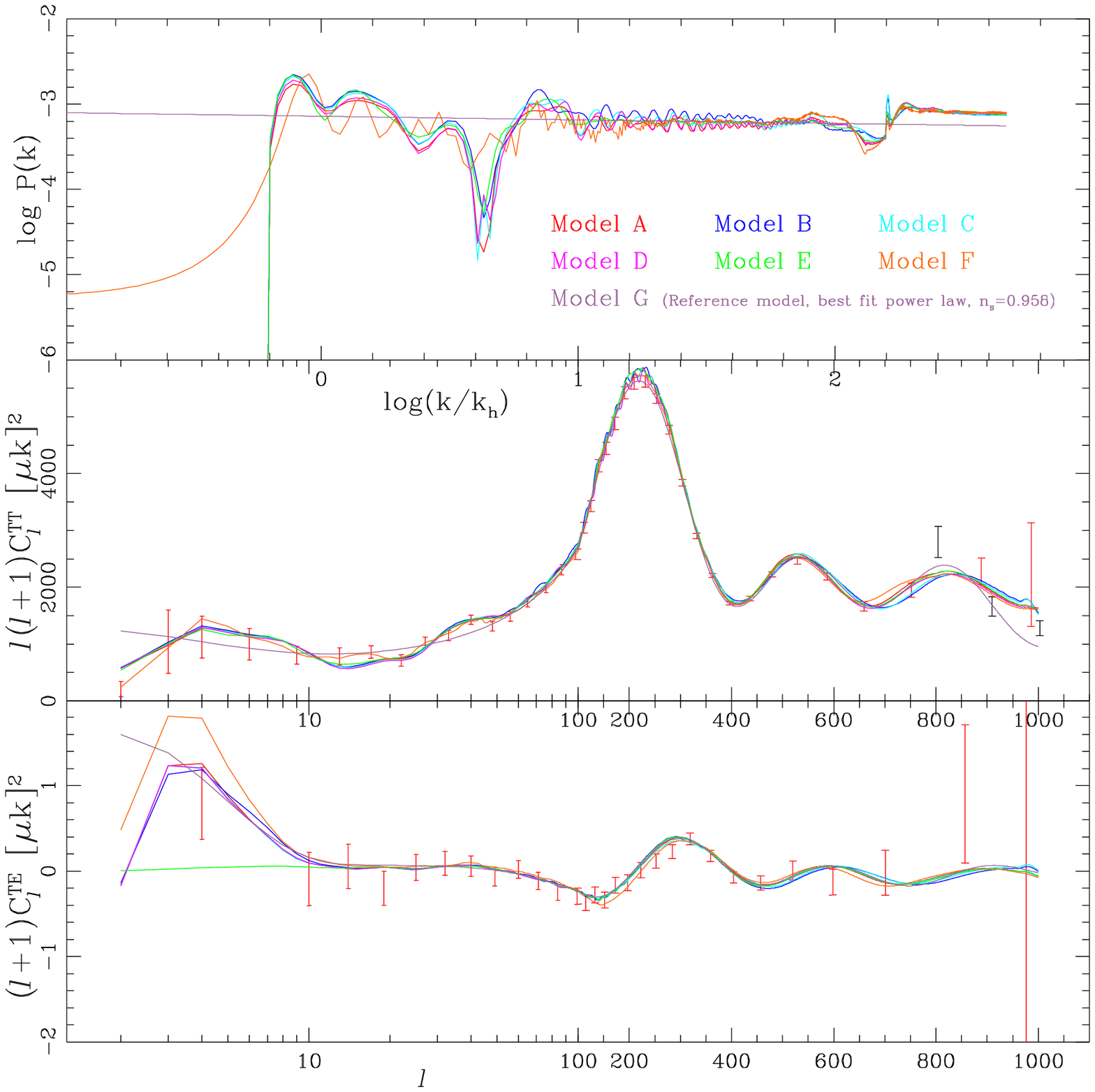}
  \caption{ Reconstructed primordial spectrum (top panel) and the resultant $C_{\ell}^{TT}$ (middle panel) and $C_{\ell}^{TE}$ (lower panel) angular power spectra are plotted for $6$ different points in the parameter space assuming a flat $\Lambda$CDM cosmological model. Cosmological parameters of {\it{Model A}}: $H_0=72, \Omega_{dm}=0.246, \Omega_b=0.05, \Omega_{\Lambda}=0.704, \tau=0.06$ and the recovered results for this model gives $\Delta \chi^2_{\rm eff}=-18.76$. Cosmological parameters of {\it{Model B}}: $H_0=63, \Omega_{dm}=0.251, \Omega_b=0.041, \Omega_{\Lambda}=0.708, \tau=0.06$ and the recovered results for this model gives $\Delta \chi^2_{\rm eff}=-4.38$. Cosmological parameters of {\it{Model C}}: $H_0=68, \Omega_{dm}=0.229, \Omega_b=0.052, \Omega_{\Lambda}=0.719, \tau=0.06$ and the recovered results for this model gives $\Delta \chi^2_{\rm eff}=-2.93$. Cosmological parameters of {\it{Model D}}: $H_0=72, \Omega_{dm}=0.229, \Omega_b=0.046, \Omega_{\Lambda}=0.725, \tau=0.06$ and the recovered results for this model gives $\Delta \chi^2_{\rm eff}=-14.52$. Cosmological parameters of {\it{Model E}}: $H_0=71, \Omega_{dm}=0.226, \Omega_b=0.044, \Omega_{\Lambda}=0.730, \tau=0.0$ and the recovered results for this model gives $\Delta \chi^2_{\rm eff}=-13.40$. Cosmological parameters of {\it{Model F}}: $H_0=50, \Omega_{dm}=0.904, \Omega_b=0.096, \Omega_{\Lambda}=0.0, \tau=0.06$ and the recovered results for this model gives $\Delta \chi^2_{\rm eff}=-26.70$. {\bf {\it{Model G}} is the {\it reference model} against which all calculated $\Delta \chi^2_{\rm eff}$s are with respect to this model. This represents the best fit power law primordial spectrum in the whole parameter space.} The red error-bars in the middle and lower panels represents the binned angular power spectrum from WMAP 3 year data. The black error-bars at the middle panel at the high $\ell$, are from ACBAR experiment~\cite{acbar}. The excess of power and the bump in the recovered $P(k)$ at the high $k$ ($log k/k_h \approx $), seems to be related to the higher measurements of the angular power spectrum at high $\ell$'s in WMAP 3 year data in comparison with the other experiments such as ACBAR.}
\label{fig_main}
\end{figure}

% show the raw deconvolved power spectrum (COLOR) and

\section{primordial power spectrum from WMAP-3 year data}

One of the most challenging questions of modern cosmology is to
find an inflationary scenario that satisfies all the cosmological
observations. The shape of the primordial power of scalar (density) and tensor (gravitational wave) generated during inflation are the key observables in this investigation. In our previous paper \cite{prd04} we introduced the Richardson-Lucy deconvolution algorithm to find the shape of the
primordial power spectrum using the cosmic microwave background data for a single point in the cosmological parameter space.
In this section we reconstruct the primordial power spectrum optimized to get the best likelihood for six different sample points in the cosmological parameter space where each of these points has a special importance. We assume a flat $\Lambda$CDM model and the differences between these $6$ cases are just in the values of the background cosmological parameters within this sub-space of parameters. In the next section we use our automated routine to perform the cosmological parameter estimation and explore a coarsely sampled but reasonable sized contiguous volume of the parameter space. In this paper, we chose to compare all the reconstructed results with the best fit power law form of the primordial spectrum in the whole parameter space, referred to as ``model G'' (see Sec. G),  rather than comparison with the result from the power law form of the spectrum for the same point in the parameter space. This highlights the effect of assuming the free form of the primordial spectrum and emphasizes on the significant improvement of the global likelihood.

\begin{table} 
\caption{Different points in the parameter space and the resultant effective likelihood from the reconstructed primordial spectrum using WMAP 3 year data. The $\Delta {\chi_{\rm eff}^2}$ is twice the logarithm of the relative likelihood with respect to the best result in the whole parameter space by assuming power law form of the primordial spectrum.}
\begin{center}
\begin{ruledtabular}
\begin{tabular}{lcccccccc}
Model & $H_0$ & $\Omega_{dm} $ & $\Omega_{b}$ & $\Omega_{\Lambda}$ & $\tau$ & $\Delta {\chi_{\rm eff}^2}$ \\
\hline
\\
Model A (compatible with SDSS) & 72.0 & 0.246 & 0.050 & 0.704 & 0.06 & -18.76 & \\ \\
Model B (compatible with 2df)  & 63.0 & 0.251 & 0.041 & 0.708 & 0.06 &  -4.38 & \\ \\
Model C (compatible with BAO)  & 68.0 & 0.229 & 0.052 & 0.719 & 0.06 &  -2.93 & \\ \\
Model D (compatible with SN Ia + BAO)  & 72.0 & 0.229 & 0.046 & 0.725 & 0.06 & -14.52 & \\ \\ 
Model E (compare to WMAP1)  & 71.0 & 0.226 & 0.044 & 0.730 & 0.0  & -13.40 & \\ \\
Model F (compatible with flat CDM)  & 50.0 & 0.904 & 0.096 & 0.0 & 0.06 &  -26.70 & \\ \\ 
\label{table}
\end{tabular}
\end{ruledtabular}
\end{center}
\end{table}

\subsection{Cosmological parameters from SDSS}
%tau   13590.0590202463

In this case we consider a flat $\Lambda$CDM cosmological model with 
cosmological parameters motivated by, and consistent with, the results of 
large scale structure observations from Sloan Digital Sky Survey (SDSS) 
\cite{SDSS03}. We use $h=0.72$ (Hubble parameter), $\Omega_{dm} =0.246$ 
(dark matter density), $\Omega_{b} =0.050$ (baryonic matter density), 
$\Omega_{\Lambda}=0.704$ ($\Lambda$ energy density) and $\tau = 0.06$ 
(optical depth). These parameters are consistent with the best fit 
results from SDSS with $h\Omega_m = 0.213 \pm 0.023$ and $\Omega_b / \Omega_m =0.17$ ( where $\Omega_m =\Omega_{dm}+\Omega_b$) for a flat $\Lambda$CDM cosmological model. We have chosen 
$\tau = 0.06$ throughout the paper, as it is one of the most reliable values for the optical depth at the present from observations of Lyman-$\alpha$ forest~\cite{lyman,lyman2}. We reconstruct the primordial power spectrum for this point in the parameter space. The reconstructed result for $P(k)$ and the resultant $C_{\ell}^{TT}$ and $C_{\ell}^{TE}$ are shown in figure \ref{fig_main} (Model A). The resultant $C_{\ell}$ (including TT and TE polarization spectra) for this point in the parameter space given by the reconstructed primordial power spectrum, can improve the effective likelihood by $\Delta \chi^2_{\rm eff}=-18.67$ with respect to the reference likelihood of model G. 

\subsection{Cosmological parameters from 2DF Galaxy Redshift Survey}
%tau   15570.8616601966
In this case we choose parameters consistent with the results from 2df galaxy redshift survey \cite{2df}.  We use $h=0.63$, $\Omega_{dm} =0.251$, $\Omega_{b} =0.041$, $\Omega_{\Lambda}=0.708$ and $\tau = 0.06$. These parameters are consistent with the results from 2df with $h\Omega_m = 0.168 \pm 0.016$ and $\Omega_b / \Omega_m =0.185 \pm 0.046$ for a flat $\Lambda$CDM cosmological model. However here we have used a marginally bigger value of matter density and marginally lower value of Hubble parameter in compare with the best fit result from 2df, but still these parameters are consistent with the 2df constraints within $1\sigma$. The reconstructed result for $P(k)$ and the resultant $C_l^{TT}$ and $C_l^{TE}$ are shown in figure \ref{fig_main} (Model B). The resultant $C_l$ for this point in the parameter space, can improve the effective likelihood by $\Delta \chi^2_{\rm eff}=-4.38$ with respect to the reference likelihood of model G.

%best result for the power law form of the primordial spectrum in the whole parameter space.      
 
\subsection{Cosmological parameters from the results of detection of baryon acoustic peak oscillations}

%tau   14668.1409796094

In this case we consider a flat $\Lambda$CDM cosmological model with cosmological parameters consistent with the results of measurements of the baryon acoustic oscillations(BAO) \cite{BAO}. We use\ $h=0.68$, $\Omega_{dm} =0.229$, $\Omega_{b} =0.052$, $\Omega_{\Lambda}=0.719$ and $\tau = 0.06$. These parameters are consistent with the best fit results from BAO with $\Omega_m h^2 = 0.130 \pm 0.011$ and $\Omega_b h^2 = 0.024$ for a flat $\Lambda$CDM cosmological model.
 The reconstructed result for $P(k)$ and the resultant $C_{\ell}^{TT}$ and $C_{\ell}^{TE}$ are shown in figure \ref{fig_main} (Model C). The resultant $C_{\ell}$ for this point in the parameter space given by the reconstructed primordial power spectrum, can improve the effective likelihood by $\Delta \chi^2_{\rm eff}=-2.93$ with respect to the reference likelihood of model G.

%best result for the power law form of the primordial spectrum in the whole parameter space.

\subsection{Cosmological parameters from observational constraints on the matter density using SN Ia and BAO data}
%tau   13986.7847868173

Model independent estimation of the matter density by using supernovae \cite{SNLS,gold} and BAO \cite{BAO} data by \cite{Shafieloo07} is used in this section to set the cosmological parameters. The total matter density, independent of the model of dark energy is found to be $\Omega_m = 0.276 \pm 0.023$. This value is, in fact, the total sum of dark matter density and baryonic matter density. We can also choose the value of baryon density in a way to be consistent with the prediction of big bang nucleosynthesis where $\Omega_b h^2 \approx 0.02$~\cite{nucleo_subir,nucleo}. We use $h=0.72$, $\Omega_{dm} =0.229$, $\Omega_{b} =0.046$, $\Omega_{\Lambda}=0.725$ and $\tau = 0.06$. These parameters are consistent with the two constraints mentioned above.  The resultant $C_l$ for this point in the parameter space driven by the reconstructed primordial power spectrum, can improve the effective likelihood by $\Delta \chi^2_{\rm eff}=-14.52$ with respect to the reference likelihood of model G. The reconstructed result for $P(k)$ and the resultant $C_{\ell}^{TT}$ and $C_{\ell}^{TE}$ are shown in figure \ref{fig_main} (Model D).

\subsection{Cosmological parameters from comparison with the results from WMAP 1 data}
%tau   14282.5435152466

In this case we use the same parameters as we used before in our previous paper \cite{prd04} where we introduced the improved Richardson-Lucy method and analyzed the WMAP 1 data. This is for comparison between our results from WMAP 1 and WMAP 3 years data. As in our previous paper, here also we use $h=0.71$, $\Omega_{dm} =0.226$, $\Omega_{bm} =0.044$, $\Omega_{\Lambda}=0.730$ and $\tau = 0.0$ for the parameters of our background cosmology. The reconstructed primordial power spectrum for this point in the parameter space, can improve the effective likelihood by $\Delta \chi^2_{\rm eff}=-13.40$ with respect to the reference likelihood of model G. This result is consistent with the result by using WMAP 1 data and there is no significant difference in features of the reconstructed results\cite{prd07}. The reconstructed result for $P(k)$ and the resultant $C_{\ell}^{TT}$ and $C_{\ell}^{TE}$ are shown in figure \ref{fig_main} (Model E).

\subsection{Cosmological parameters from Standard Cold Dark Matter (SCDM) model}
% tau   11837.7001092157
In this case we assume a Cold Dark Matter universe (CDM) where the energy density of the dark energy is assumed to be zero. For $h=0.50$, $\Omega_{dm} =0.904$, $\Omega_{b} =0.096$, $\Omega_{\Lambda}=0.0$ and $\tau = 0.06$, we could improve the effective likelihood by $\Delta \chi^2_{\rm eff}=-26.70$ which shows that by assuming a free form of the primordial spectrum, the standard CDM model of a flat universe can be very well fitted to the CMB data alone. Studies by \cite{huntsarkar07} have shown that a CHDM model of the universe (which there is also no dark energy) also can have a good fit to the WMAP 3 years data. It is very interesting that for this point in the parameter space we could get a big improvement in the effective likelihood. We should note here that for this point in the parameter space, we have set $\Omega_b h^2 = 0.024$ which is in agreement with big bang nucleosynthesis however this point in the parameter space is not well fitted with other cosmological observations, like large scale structure observations or the supernovae data. Results are shown in figure \ref{fig_main} (Model F).

\subsection{Reference model: cosmological parameters from best fit power law to the WMAP 3 year data}
%tau   14555.5918280376
    
Interestingly, for this point in the parameter space, we could not significantly improve the effective likelihood by considering the free form of the primordial spectrum where we used $h=0.732$, $\Omega_{dm} =0.1967$, $\Omega_{b} =0.0416$, $\Omega_{\Lambda}=0.7617$ and $\tau = 0.089$.  In fact adding features to the form of primordial spectrum for this point in the parameter space could not improve the resultant effective likelihood. We interpret this to arise from the fact the cosmological parameters themselves adjust with a large number of degrees of freedom to the fit comparable to the freedom in the primordial power spectrum (encoded in a finite number of wavelet coefficients). In short, all the cosmological parameters have been chosen in way to give us the best likelihood by strong assumption of power law form of the primordial spectrum. {\bf In this paper, the resultant $\chi^2_{\rm eff}$ of different models are compared with this model as the reference model.}

Different assumed models with their parameters and their resultant likelihoods are shown in table \ref{table}.

\begin{figure*} 
\centering
\begin{center} 
\vspace{-0.05in}
\centerline{\mbox{\hspace{0.in} \hspace{2.1in}  \hspace{2.1in} }}
$\begin{array}{@{\hspace{-0.1in}}c@{\hspace{0.4in}}c@{\hspace{0.3in}}c}
\multicolumn{1}{l}{\mbox{}} &
\multicolumn{1}{l}{\mbox{}} &
\multicolumn{1}{l}{\mbox{}} \\ [-0.5cm]
 
\includegraphics[scale=0.45, angle=0]{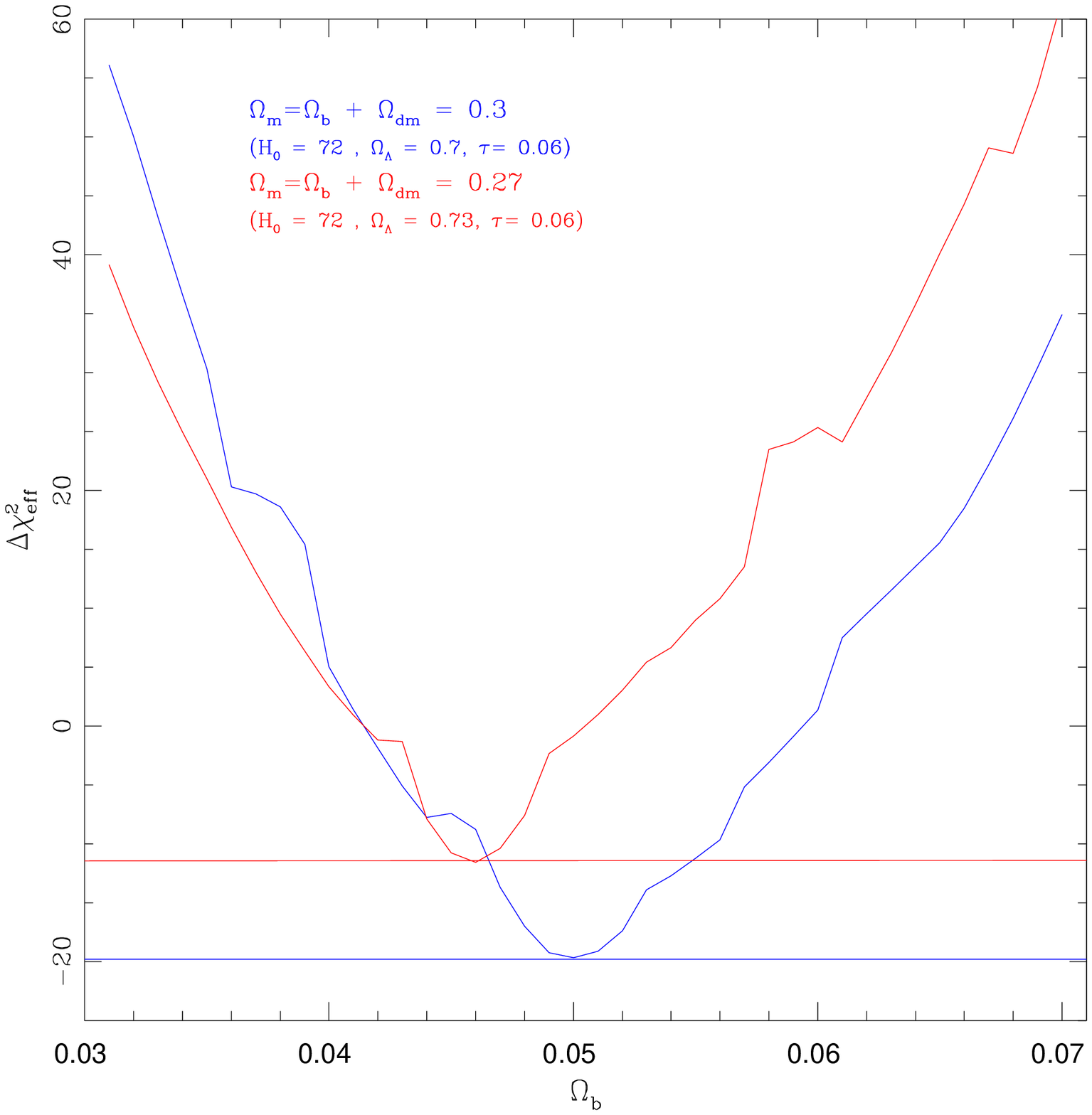}
 
\includegraphics[scale=0.45, angle=0]{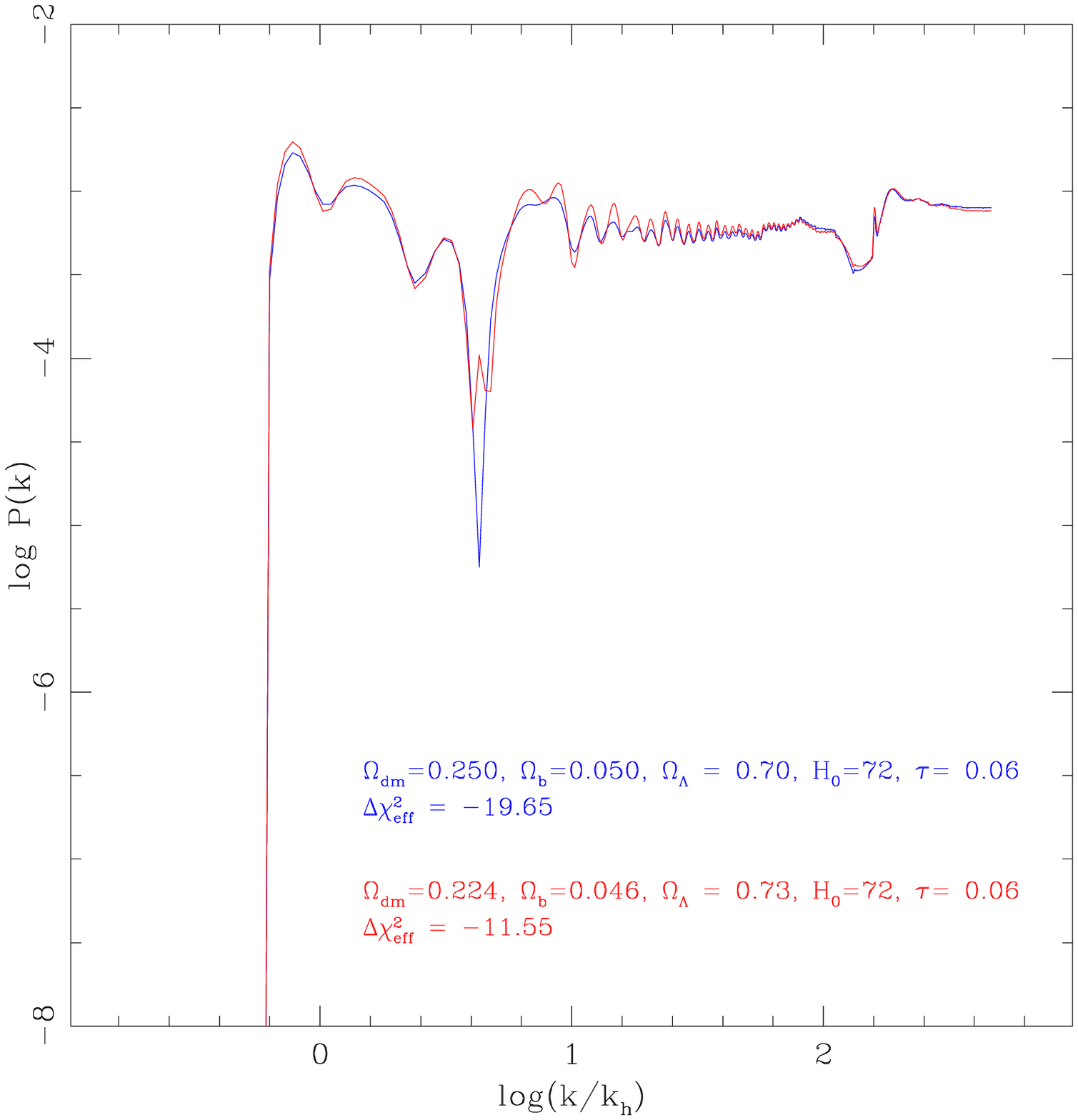}
%\mbox{\bf (a)}
\end{array}$

\end{center}
\caption{ A 1-D slice ($\Omega_m =constant$) through the cosmological parameter space demonstrates that the data retains strong discriminatory power in the cosmological parameter space even when there is full freedom in choosing the primordial power spectrum. {\bf Left panel:} Plot of $\Delta \chi^2_{\rm eff}$ of the reconstructed results with respect to the reference likelihood of model G, by assuming free form of the primordial spectrum, for a flat $\Lambda$CDM model with $h_0 = 0.72$, $\tau = 0.06$ and $\Omega_{\Lambda} =0.70$ and  $\Omega_m= \Omega_{b} + \Omega_{dm} = 0.30$ for different values of $\Omega_{b}$ (blue line). The red curve is for similar models except for $\Omega_m= \Omega_{b} + \Omega_{dm} = 0.27$. Clearly, `optimizing' over the primordial power spectrum allows us to get significantly higher likelihood ($\Delta \chi^2=-19.65$) for  $\Omega_m = 0.30$ compared to $\Omega_m = 0.27$ ($\Delta \chi^2=-11.55$). This demonstrates that even though we allow a free form of the primordial spectrum, the data does show very strong preference for particular values of cosmological parameters. {\bf Right panel:} Reconstructed primordial spectrum, $P(k)$, for a flat $\Lambda$CDM model with $ \Omega_{b}=0.050, \Omega_{dm}=0.25, h_0 = 0.72$, $\tau = 0.06$(blue line). For these parameters of $ \Omega_{b}$ and $\Omega_{dm}$, we could get the best likelihood for $\Omega_m = 0.30$. The red line is the reconstructed $P(k)$ for a flat $\Lambda$CDM model with $ \Omega_{b}=0.0460, \Omega_{dm}=0.224, h_0 = 0.72$, $\tau = 0.06$. For these parameters of $ \Omega_{b}$ and $\Omega_{dm}$, we could get the best likelihood for the $\Omega_m = 0.27$. It is clear that the reconstructed $P(k)$ for these two points in the cosmological parameter space are very similar. However the resultant $\Delta \chi^2_{\rm eff}$ for these two points in the parameter space shows a big difference.}
\label{fig_baryon}
\end{figure*}

\section{Toward Cosmological Parameter Estimation}

It is very important to note that despite of allowing a free form for the primordial spectrum, not all cosmological models (i.e., all points in the parameter space) can be fitted to the data equally well. We clearly show that some points in the cosmological parameter space fit the WMAP 3 year CMB data better than the other points, by 'optimizing' the likelihood over a free form of the primordial spectrum. We conjecture that the positive definiteness of the primordial spectrum does not allow us to fit all the points in the parameter space to the data equally well, and some points will have a better fit to the data.
% The final goal of this program of research is to carry out cosmological parameter estimation after optimizing the likelihood  at any point over the primordial spectrum using our automated method. However, at this time this is beyond our computational resources and we defer this to future publication. 
In this section we would like to present strong evidence that despite of allowing a free form of the spectrum, the derived likelihoods do strongly discriminate between different neighboring points in the parameter space. As an example, for a flat $\Lambda$CDM model, we fix the values of $H_0 = 72$, $\tau = 0.06$ and $\Omega_{\Lambda} =0.70$ and we vary the values of $\Omega_{b}$ and $\Omega_{dm}$ keeping the total fixed at $\Omega_m= \Omega_{b} + \Omega_{dm} = 0.30$ and calculate the likelihood. We find a minimum in the value of the $\chi^2_{\rm eff}$ around  $\Omega_{b} = 0.050 $ and $\Omega_{dm} = 0.250$ which shows that the data prefers this combination among the models with $\Omega_m=0.30$. 

In parameter estimation, other cosmological observations can be used to put constraints on the parameter space. For example, some region of the parameter space may be in agreement with CMB data, but being ruled out strongly by other observations. However in our example, the best result seems to be well in agreement with all other cosmological observations, including large scale structure observations from SDSS \cite{SDSS03}, supernovae data by SNLS \cite{SNLS} and Gold \cite{gold}, detection of baryon oscillations \cite{BAO} and it is also in agreement with big bang nucleosynthesis. In figure \ref{fig_baryon} we see the resultant  $\Delta \chi^2_{\rm eff}$ versus different values of baryonic matter (left panel-blue line) and the reconstructed $P(k)$ for the best combinations of $\Omega_{b}$ and $\Omega_{dm}$ assuming $\Omega_m = 0.30$ (right panel-blue line). The red curves in figure \ref{fig_baryon} have the same characteristic but $\Omega_{\Lambda} =0.73$ and $\Omega_m = 0.27$. Clearly, `optimizing' over the primordial power spectrum allows us to get  significantly higher likelihood ($\Delta \chi^2=-19.65$) for  $\Omega_m = 0.30$ compared to $\Omega_m = 0.27$ ($\Delta \chi^2=-11.55$). 

\begin{figure}[h]
  \includegraphics[scale=0.6, angle=0]{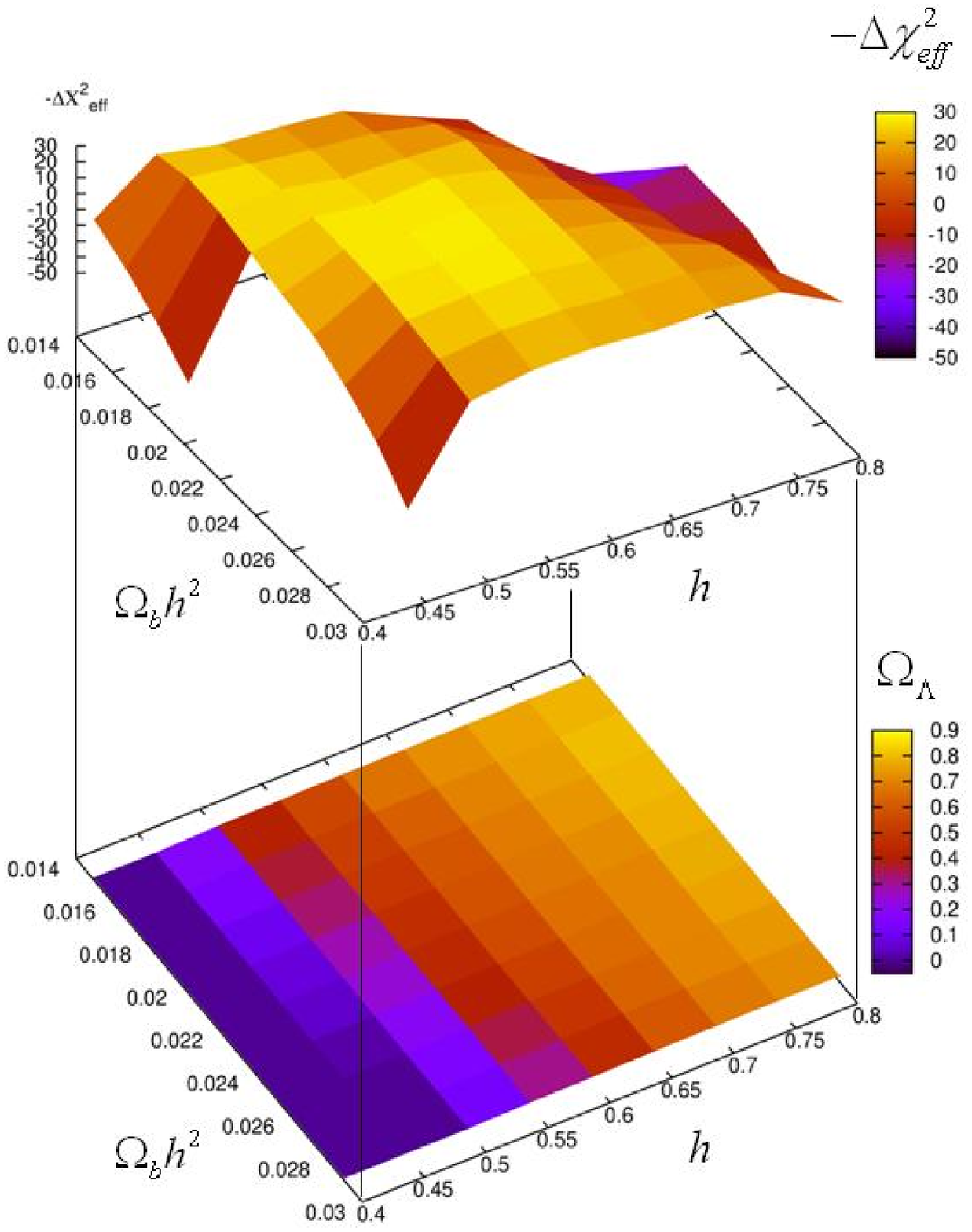}
  \caption{ A coarse resolution and limited volume exploration of the cosmological parameter space demonstrates that the data retains strong discriminatory power in the cosmological parameter space even when there is full freedom in choosing the primordial power spectrum. The resultant  $-\Delta \chi^2_{\rm eff}$ is shown (in Z axis and also in color indicated by the tool bar in the upper panel) versus different values of Hubble parameter (X axis in both upper and lower panel) and $\Omega_b h^2$ (Y axis in both upper and lower panel). The lower panel shows the relative values of the $\Omega_{\Lambda}$ in our parameter space (indicated by color in the lower panel). We have assumed here $\tau=0.06$. }
\label{fig_param}
\end{figure}

Here we have only considered two limited 1-D slices in the cosmological parameter space, but our main aim is to do the cosmological parameter estimation for the whole volume of the parameter space. However as it has been mentioned before, doing a cosmological parameter estimation in the whole parameter space would be computationally very expensive. 
Recently few new methods of parameter estimation have been proposed which are claimed to be much faster than the usual methods of Monte Carlo Markov Chain or grid sampling. These new methods may be suitable for our purpose but they are still applicable for a one dimensional space and they need to be modified to be applied in our problem\cite{kat,will,amir,hobson}. 
Though it is still difficult and beyond our abilities to do the cosmological parameter estimation in the whole parameter space and with a high resolution, still we can do it for a reasonably large contiguous volume of the parameter space. In this part, we present the results for the cosmological parameter estimation by fixing the value of $\tau=0.06$ (which we have chosen throughout the paper) and varying the values of $\Omega_b$, $\Omega_{dm}$ and $h$ in a large volume of the parameter space. In our analysis we use the following priors: $ 35 < H_0 < 85$, $ 0.012 < \Omega_b h^2 < 0.030 $, $ 0.04 < \frac{\Omega_b}{\Omega_{dm}} < 0.30 $ and $ 0 < \Omega_{\Lambda} < 1 $. Any of the given priors for $\Omega_b$, $\Omega_{dm}$ and $h$ has been divided by eight equally-spaced points whose combinations will generate our assumed initial sampling in the parameter space. $\Omega_{\Lambda}$ is derived from the other parameters assuming a flat universe.  Our motivation to choose these wide priors are from the strong limits from other astronomical and cosmological observations. In figure \ref{fig_param} we see the resultant  $-\Delta \chi^2_{\rm eff}$ (in Z axis and also in color indicated by the tool bar in the upper panel) versus different values of Hubble parameter (X axis in both upper and lower panel) and $\Omega_b h^2$ (Y axis in both upper and lower panel). The lower panel shows the relative values of the $\Omega_{\Lambda}$ in our parameter space (indicated by color in the lower panel).
% We can see that the WMAP 3 years data clearly prefers some points in the parameter space to the points if we assume a free form of the primordial spectrum. 

The best likelihood has been derived for $\Omega_b = 0.084$, $\Omega_{dm} = 0.764$, $\Omega_{\Lambda} = 0.152$ and $H_0 = 50$ with $\Delta \chi^2_{\rm eff} = -29.282$ with respect to the reference likelihood of model G. Another point in the parameter space with $\Omega_b = 0.058$, $\Omega_{dm} = 0.416$, $\Omega_{\Lambda} = 0.526$ and $H_0 = 60$, which is more compatible with the other cosmological observations, also has a very good optimized likelihood with $\Delta \chi^2_{\rm eff} = -29.014$ with respect to the reference likelihood of model G, which is very close to our best result.

%We can see that the data prefer  $ 50 \leq  H_0 \leq 60$ and  $ 0.020 \leq \Omega_b h^2 \leq 0.024 $ by assuming a free form of the primordial spectrum. 

Results also show that for a very wide range of $\Omega_{\Lambda}$, even when its density is close to zero, we can have a very good fit to the WMAP 3 year data which is expected, as CMB data alone are not very much sensitive to the $\Lambda$ density.

%It is obvious that we need to do the higher resolution of the parameter estimation and probably on a bigger volume of the parameter space to get a better result. 
Due to computational limitations, we could not do a fine resolution cosmological parameter estimation. Work is still in progress and results will be published as soon as we accomplish the computations with the WMAP 5 year data.  
 
%This motivates us to work towards precise cosmological parameter estimation, allowing full freedom to the form of the primordial spectrum. 

We hope that this would provide a completely different perspective to cosmological parameter estimation unbiased by prejudices of the early universe.

\section{Conclusion}

We present the reconstruction of an 'optimal' primordial
power spectrum for flat $\Lambda$CDM cosmological models for different sample points in the parameter space. 

In section IV we have chosen these sample of points to be consistent with different independent cosmological observations, or consistent with special theoretical models. Almost in all cases the recovered spectrums improves the resultant effective likelihood significantly in comparison with the best fit power law form of the primordial spectrum in the whole parameter space. 
In section V we generalized our study to a much bigger sample of points where we performed the cosmological parameter estimation in a large volume of the parameter space.

There are some important conclusions that can be drawn from our results. Though the published results from WMAP team after release of WMAP 3 year data suggests the lower value of matter density in comparison with the other cosmological observations like large scale structure observations from SDSS, supernovae and detection of baryon acoustic peak oscillations, our analysis shows that by assuming a free form of the primordial spectrum, CMB data can be well fitted to the models with the higher value of matter density compatible with the other cosmological observations. Our preliminary studies in the parameter space, show some evidence that by assuming the free form of the primordial spectrum models with higher value of matter density are better fitted to the WMAP 3 year data than models with low value of matter density. Another important result is that a standard CDM model of the universe, can also be very well fitted to the WMAP 3 year data by assuming the free form of the primordial spectrum. In fact for a SCDM model, we could get one of our best recovered results by improving the likelihood around $\Delta \chi^2_{\rm eff}=-27$ with respect to the best result for the power law form of the spectrum in the whole parameter space(reference likelihood of model G). 
The features of the recovered results for all these points in the parameter space have something in common. They all show a sharp cut of around the horizon and some significant features after the horizon scale. It has been shown before by \cite{prd07} that the effect of these features around the horizon are very important in improving the likelihood. These results are in agreement with the results from our previous paper where we studied the WMAP 1 data. 

Assuming a free form of the primordial spectrum, we derived the best likelihood for a $\Lambda$CDM model at $\Omega_b = 0.084$, $\Omega_{dm} = 0.764$, $\Omega_{\Lambda} = 0.152$ and $H_0 = 50$ within the explored region with $\Delta \chi^2_{\rm eff} = -29.282$ with respect to the reference likelihood of model G. Another point in the parameter space with $\Omega_b = 0.058$, $\Omega_{dm} = 0.416$, $\Omega_{\Lambda} = 0.526$ and $H_0 = 60$, which is more compatible with the other cosmological observations, also has a very good optimized likelihood with $\Delta \chi^2_{\rm eff} = -29.014$ with respect to the reference likelihood of model G.

%with  $ 50 \leq  H_0 \leq 60$, $ 0.020 \leq \Omega_b h^2 \leq 0.024 $ and  $ 0.0 \leq  \Omega_{\Lambda} \leq 0.7$. 

The differences between our results in parameter estimation and the results from WMAP team assuming a power law form of the primordial spectrum motivates us to work towards precise cosmological parameter estimation allowing full freedom to the form of the primordial spectrum, with a higher resolution in the spacing of the parameters and also considering the whole parameter space.

%The next step of our research program is to carry out a ***** high resolution*****cosmological parameter estimation in the whole parameter space by assuming the free form of the primordial spectrum. This work is limited by the computational cost (but within the scope of existing computational facilities world-wide) and the ****complete**** results would be published later in a future communication. 
In this work, we use the modestly determined quality of CMB polarization spectra from WMAP (TE and EE) simply as a consistency check. However, CMB polarization (EE) spectra from Planck Surveyor is expected to be good enough to allow us to extend our deconvolution method simultaneously to CMB temperature and polarization.

\section{Acknowledgment}
We acknowledge the use of the Legacy Archive for Microwave Background Data Analysis (LAMBDA).  Support for LAMBDA is provided by the NASA Office of Space Science. We acknowledge use of HPC facilities at IUCAA.

\newpage
\newpage

\end{document}